# Can Connected Autonomous Vehicles really improve mixed traffic efficiency in realistic scenarios?

Mohit Garg, Cian Johnston and Mélanie Bouroche

*Abstract*— Connected autonomous vehicles (CAVs) can supplement the information from their own sensors with information from surrounding CAVs for decision making and control. This has the potential to improve traffic efficiency. CAVs face additional challenges in their driving, however, when they interact with human-driven vehicles (HDVs) in mixed-traffic environments due to the uncertainty in human's driving behavior e.g. larger reaction times, perception errors, etc. While a lot of research has investigated the impact of CAVs on traffic safety and efficiency at different penetration rates, all have assumed either perfect communication or very simple scenarios with imperfect communication. In practice, the presence of communication delays and packet losses means that CAVs might receive only partial information from surrounding vehicles, and this can have detrimental effects on their performance. This paper investigates the impact of CAVs on traffic efficiency in realistic communication and road network scenarios (i.e. imperfect communication and large-scale road network). We analyze the effect of unreliable communication links on CAVs operation in mixed traffic with various penetration rates and evaluate traffic performance in congested traffic scenarios on a large-scale road network (the M50 motorway, in Ireland). Results show that CAVs can significantly improve traffic efficiency in congested traffic scenarios at high penetration rates. The scale of the improvement depends on communication reliability, with a packet drop rate of 70% leading to an increase in traffic congestion by 28.7% and 11.88% at 40% and 70% penetration rates respectively compared to perfect communication.

## I. INTRODUCTION

Human-driven vehicles produce stop-and-go waves i.e. high speed variations, due to large reaction times and perception errors. This stop-and-go behavior is the main cause of traffic congestion, resulting in reduced traffic safety and efficiency [8]. CAVs have the potential to dampen these waves and thereby improve traffic efficiency without compromising traffic safety by adopting a suitable car-following control strategy [11]. Adaptive cruise control (ACC) and cooperative adaptive cruise control (CACC) are two major car-following control strategies employed in CAVs for their autonomous driving behavior in the longitudinal direction. ACC exploits autonomous vehicles' faster reaction time to reduce waves, while cooperative adaptive cruise control allows vehicles to follow their preceding vehicles with a shorter time headway by enabling the ego vehicle to receive information (e.g. position, speed, and acceleration) from its preceding vehicle through wireless communication links [17]. CACC controller performance is generally measured using the concept of string stability. String stability ensures that fluctuations in leading vehicles behavior (e.g. acceleration/deceleration due to scenarios such as Stop and Go, Hard Brake, Cut-In, Cut-Out, human distractions, etc.) does not propagate downstreams into the string [12].

As connected and autonomous vehicles will appear on roads gradually, they will coexist with human-driven vehicles for a long time. During this time, the traffic is likely to be a mix of different kinds of connected and autonomous vehicles (such as ACC-equipped vehicles, CACC-equipped vehicles, connected vehicles with level 4 automation technology) as CAV technology progresses [4]. Many studies have explored the impact of CAVs on traffic performance. These show that CAVs can significantly improve traffic efficiency at high penetration rates by leveraging the CACC technology, however, a detrimental effect is observed at low penetration rates [11].

These studies have assumed reliable inter-vehicle communication. In practice, however, the information exchange is likely to suffer from packet losses or message reception delays due to limited transmission bandwidth, channel fading and interference [17]. When a vehicle cannot establish a link with the previous vehicle, it degrades its mode to ACC, resulting in reduced traffic efficiency due to the increased time headway in the ACC mode [3]. A few researchers have analyzed the effect of unreliable communication links on CAVs performance but only in limited scenarios i.e. with a small number of vehicles. This paper simulates CAVs in realistic communication and road network scenarios with different market penetration rates, using real traffic data of an Irish motorway and then evaluates the effect of imperfect communication on different performance indices such as travel rate and relative congestion index (RCI). Consequently, the main contribution of the present study is to investigate the impact of CAVs on mixed traffic efficiency in realistic communication and mobility scenarios.

This paper is organized as follows. The next section reviews the related work exploring the impact of connected autonomous vehicles on traffic efficiency. Section III presents the simulation set up and evaluation scenarios based on the Plexe simulator. Section IV analyzes the simulation results of different scenarios and discusses them in detail. The final section concludes the presented work and discusses the future scope of the work.

## II. RELATED WORK

CAV technology is an active area of research due to its potential to improve our road transport system. A number of simulation studies have analyzed and quantified the impact

Mohit Garg, Cian Johnston and Mélanie Bouroche are with the School of Computer Science and Statistics, Trinity College Dublin, Ireland `mgarg@tcd.ie, johnstc@tcd.ie, Melanie.Bouroche@tcd.ie`

of CAVs on traffic performance (i.e. traffic safety, efficiency and traffic flow stability) at different market penetration rates [1], [2], [7], [9], [10]. Results suggest that at low penetration rates (usually less than 25%) CAVs would have a negative impact on mixed traffic efficiency, while at medium-to-high penetration rates (usually 40% or more), they would improve it significantly. While most studies focus on analyzing the impact of CAVs using unrealistic traffic mobility scenarios i.e. simple road networks and traffic demand, Gueriau and Dusparic investigated traffic performance using realistic road network and traffic data (i.e. a large-scale road network with real traffic data of the M50 motorway, in Ireland) [7]. Results confirm that at low penetration rates, traffic efficiency is affected negatively. These results, however, are very sensitive to the car-following models chosen to describe driving behaviour [16]. Early studies showed that when time headway settings and car-following models were chosen according to field experiments results, CAVs can only improve traffic efficiency at medium-to-high penetration rates [1], [2], [4], [13]. Furthermore, recent studies show that CAVs can further improve traffic efficiency when their car-following model parameters are calibrated and then optimized based on real traffic data [9]. It is also being reported that effects of CAVs are more pronounced in congested traffic scenarios rather than free flow [2].

The aforementioned studies, however, assume perfect communication. In practice, communication delays and packet drops might affect the results. Very few studies have analyzed the impact of CAVs on traffic performance in the presence of communication impairments i.e. delays and packet losses [3], [8]. The study presented in [8] investigated how CAVs and connected human-driven vehicles, which are organized in the form of a string (of three, eight and twenty vehicles), affect traffic congestion in the presence of time-varying delays, and showed that the presence of communication impairments has a detrimental effect on CAVs performance because their control algorithms use outdated information. Furthermore, a recent study based on a string of only three CACC-equipped vehicles highlighted that the potential benefits of CAVs are compromised in imperfect communication environments due to degradation of CACC-equipped vehicle into ACC-equipped vehicle, resulting in reduced traffic efficiency at a string stable time headway [3].

Table I summarizes the study of CAVs (of different automation levels) on traffic performance, for different penetration rates and traffic scenarios. It can be seen that different studies have chosen different time headways values, and as it plays a major role in mixed traffic efficiency, it should be chosen based on field-experiments performed in realistic scenarios. Furthermore, none of the studies have investigated the impact of CAVs in realistic communication and traffic scenarios. Most studies assume perfect communication with large-scale road networks and a few studies investigate the effect of imperfect communication links on traffic performance considering only a small number of vehicles in the string. To fill this research gap, this paper builds on top of the work by Gueriau and Dusparic [7] to investigate the impact of CAVs on traffic efficiency in realistic scenarios in terms of vehicle modeling, road network and communication.

## III. SIMULATION SET-UP AND EVALUATION SCENARIOS

### A. Platform choice

Performing traffic simulations with both realistic driving and communication model require using the integration of a microscopic vehicular traffic simulator and a communication network simulator. Veins is a popular open source framework which integrates the SUMO traffic simulator with the OMNET++ network simulator. Another option is, TraNS an another open-source integrated simulation framework (integrating the SUMO traffic simulator with the NS2/NS3 network simulator) to perform realistic vehicular networking simulations. This study uses the Plexe simulator, which is an extension of VEINS for cooperative driving applications such as CACC and platooning [14]. The choice of this simulator was partially influenced by the integration available between microscopic traffic simulator and a communication network simulator, and partially by related work [8], [14], [21].

### B. Vehicle modelling and parameters

Various car-following and lane-changing models are typically used to describe the driving behavior in longitudinal and lateral directions. In this work, the field-tested CACC and ACC models are used for modeling the car-following behavior of CAVs in CACC and ACC modes respectively [15]. Due to their calibration and validation during real experiments in different scenarios, these car-following models represent realistic speed and acceleration profiles for CAVs. The well-known IDM car-following model is used for modeling the behavior of a human-driven vehicle and the default lane-changing model in SUMO is used for modeling the lane-changing behavior of both CAVs and human-driven vehicles [18]. These particular models were chosen after reviewing related work [2], [3]. The car-following and lane-changing models parameters are presented in Table II. Different time headway settings are selected depending on whether a CAV is following a human-driven vehicle (1.1s) or a CAV (0.6s) or the vehicle is human-driven (1.5s) [15] [20]. The lcStrategic and lcCooperative parameters of the lane changing model are changed from their default value of 1 to 0.5 to better replicate the lane-changing in real life [7]. Other parameters were left at their default values [18].

### C. CACC controller

Plexe supports a variety of car-following control algorithms i.e. CACC controllers for cooperative longitudinal control of CAVs. In this work, the Ploeg CACC controller, designed using the widely-used constant-time-headway policy and the one-vehicle ahead (i.e. predecessor-following) information flow topology, in which an ego vehicle receives information (i.e. position, speed and acceleration) from only its immediate leading vehicle is employed [16], [22]. The Ploeg controller gain values are selected as advised by the authors of Plexe [14] and are reported in Table III.

TABLE I: Comparative study of CAVs operation in mixed traffic and their impact on traffic performance

| Reference | Vehicle type | MPR (%) | Time headway settings | Road network | Traffic scenario | Realistic communication | Traffic efficiency | Traffic safety |
|---|---|---|---|---|---|---|---|---|
| Shladover et al. [1] | ACC, CACC, HDVs, Connected HDVs | 10, 20, 30, 40, 50, 60, 70, 80, 90 | 1.48 to 1.8s (for HDVs), 1.1 to 2.2s (ACC mode), 0.5s (CACC mode) | Single-lane 6.5 km long highway | Over-saturated | ✗ | ✓ | ✗ |
| Arnaout and Arnaout [2] | CACC, HDVs | 0, 20, 40, 60, 80, 100 | 0.8 to 1s (for HDVs), 0.5s (CACC mode) | Four-lane 6 km long highway | Moderate, saturated & over-saturated | ✗ | ✓ | ✗ |
| Talebpour and Mahmassani [10] | CAVs, AVs, Connected HDVs | 10, 25, 50, 75, 90 | 1 to 1.5s | Single lane highway | Free flow | ✗ | ✓ | ✗ |
| Liu et al. [13] | CAVs, HDVs, Connected HDVs | 0, 20, 40, 60, 80, 100 | 11.4s (for HDVs), 0.6s (CACC mode) | Four-lane 18 km long highway | Free-flow & over-saturated | ✗ | ✓ | ✗ |
| Navas and Milanés [3] | CAVs, HDVs | - | 1.5s (for HDVs), 0.6s (CACC mode) | Single lane highway | INRIA experiment platform with three cycabs | ✓ | ✓ | ✗ |
| Ye and Yamamoto [4] | CACC, HDVs | 0, 20, 30, 40, 50, 60, 70, 80, 90 | 0.5 to 1.1s | Two-lane 10 km long highway | Free-flow | ✗ | ✓ | ✓ |
| Papadoulis et al. [5] | CAVs, HDVs | 0, 25, 50, 75, 100 | 0.6s (CACC mode) | Three-lane 44.27 km long motorway | Free-flow, saturated & over-saturated | ✗ | ✓ | ✓ |
| Vaio et al. [8] | CAVs, HDVs | - | Distance headway-20m | Single-lane | String of three, eight and twenty vehicles | ✓ | ✓ | ✗ |
| Liu and Fan [9] | CAVs, HDVs | 0, 10, 20, 30, 40, 50, 60, 70, 80, 90, 100 | - | Four-lane highway | Over-saturated | ✗ | ✓ | ✗ |
| Guériau and Dusparic [7] | CAVs, HDVs | 0, 2.5, 7, 20, 40, 70 | 1.2 to 1.5s (for HDVs), 0.6 to 0.8s (CACC mode) | Urban, national and motorway | Free-flow, saturated, over-saturated | ✗ | ✓ | ✓ |

TABLE II: Car-following and lane-changing models for simulated CAVs and HDVs

| Parameters | CAVs | HDVs |
|---|---|---|
| **Car-following model** | Field-tested CACC [15] | IDM [18] |
| Speed deviation (%) | 0.05 | 0.1 |
| Time headway (s) | 0.6 | 1.5 |
| Minimum gap (m) | 1.5 | 2.5 |
| Max accel. (m/s$^2$) | 2.9 | 1.5 |
| Deceleration (m/s$^2$) | 7.5 | 7.5 |
| Emerg. decel. (m/s$^2$) | 9 | 9 |
| **Lane-changing model** | Default (LC2013) [19] | Default (LC2013) |
| lcStrategic | 0.5 | 0.5 |
| lcCooperative | 0.5 | 0.5 |

Furthermore, Plexe allows the implementation of the upper-layer CACC control considering realistic communication and adopts a 3$^{rd}$ order dynamic model to represent realistic vehicle dynamics for the lower-level control. This realistic vehicle dynamic model comprises an actuator lag parameter to model delays and time lags in engine response, sensors and actuators.

### D. Communication network parameters

The parameters in Table III relate to the communication network parameters used by OMNeT++/Veins. Each CAV is equipped with a network interface card according to the IEEE 802.11p communication protocol. CAVs transmit information at 10 Hz frequency using the default network parameters [14]. To simulate imperfect communication, the frameErrorRate parameter is used to create artificial losses at the MAC layer. Two values of frameErrorRate i.e. 0 and 0.7 are chosen to represent no packet drops and 70% packet drops respectively.

### E. Road network scenario

A large-scale road network (the M50 motorway, in Ireland as shown in Fig. 1) with real traffic data originally created by Gueriau and Dusparic[1], is used to perform mixed traffic simulations in realistic scenarios [7]. This road network consists of a large number of vehicles (up to 25,316 vehicles during the busiest time period, 07:00-08:00) on a 7-km 4-lane stretch of the M50 motorway Road. We choose a simulation time window of 30 minutes i.e. 07:00-07:30 to reduce simulation overhead. When performing a simulation to analyze the impact in a particular time period, a simulation parameter manager.firstStepAt is leveraged to advance the simulation to a specified point in time without the additional overhead of network simulation, and then proceed to perform the rest of the network simulation logic with a fully loaded traffic network. The Plexe scenario with related simulation files of this work is available in the git repository[2].

TABLE III: Vehicle model and communication related parameters

| Vehicle model parameters | |
|---|---|
| Parameter | Value |
| Vehicle length | 5m |
| Longitudinal dynamics | $3^{rd}$ order |
| Actuator lag | 0.5s |
| Controller gains | 0.2 (distance gain), 0.7 (speed gain) |
| **Communication parameters** | |
| Communication protocol | IEEE 802.11p |
| Channel data rate | 6 Mbps |
| Beacon frequency | 10 Hz |
| Beacon size | 200 bytes |
| Packet losses | Bernouli loss model |
| Transmission power | 20 dBm |
| Antenna type | Monopole |
| Analogue Model | SimplePathLossModel with thresholding |

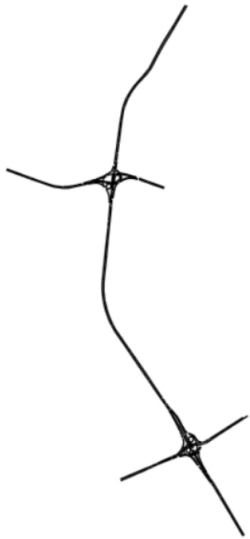

Fig. 1: Rendering of the M50 motorway road network

[1] https://github.com/maxime-gueriau/ITSC2020_CAV_impact/tree/master/Motorway
[2] https://github.com/gargmohit24/ITSC_2021

## IV. RESULTS AND DISCUSSIONS

The deployment of CAVs is expected to improve traffic efficiency, though it is unclear to what extent this has been evaluated in realistic scenarios. In this paper, a number of simulation experiments were conducted to evaluate traffic performance at different penetration rates (i.e. 0%, 20%, 40%, 70%) with and without packet drops (i.e. 0 and 0.7 packet error rate). This resulted in one baseline scenario with no CAVs, three scenarios with CAVs and no packet losses and three scenarios with CAVs and packet error rate of 0.7, as reported in table IV. Each of these 7 scenarios is evaluated in terms of traffic efficiency.

TABLE IV: Summary of traffic scenarios used for experiments

| Traffic hours [7] | CACC controller [22] | MPR (%) | PER (%) |
|---|---|---|---|
| Congested (07:00-07:30) | NA | 0 | NA |
| Congested (07:00-07:30) | Ploeg | 20 | 0 |
| Congested (07:00-07:30) | Ploeg | 20 | 70 |
| Congested (07:00-07:30) | Ploeg | 40 | 0 |
| Congested (07:00-07:30) | Ploeg | 40 | 70 |
| Congested (07:00-07:30) | Ploeg | 70 | 0 |
| Congested (07:00-07:30) | Ploeg | 70 | 70 |

### A. Key performance indicators

A considerable number of studies have investigated mixed traffic efficiency using travel rate and congestion index as key performance indicators. Travel rate gives the information about how much time a vehicle takes to travel over a certain edge of road network, while relative congestion index (RCI) advises about the state of traffic flow operations more accurately. An RCI value of greater than 2 indicates a very high traffic congestion. A study in [23] revealed that the travel rate cannot explicitly indicate the state of traffic flow for congested traffic scenarios. In congested traffic scenarios, a vehicle either travels at a low speed with almost no speed variations or a vehicle travels at a slightly high speed with many instances of speed variations. In this paper, we therefore measure both travel rate and relative congestion index on each edge of the simulated road network.

### B. Simulation results

Fig. 2 and 3 show the simulation results of the travel rate and congestion index for the case of only human-driven vehicles (0% penetration rate of CAV). Results indicate that both travel rate and relative congestion index are high in vast parts of the road network.

Fig. 4 shows the travel rate at different penetration rates (20%, 40%, 70%), with and without packet drops. From Fig. 4a and 4b, it is observed that, as in other studies, at a low penetration rate (20%), CAVs worsen the traffic flow conditions near the south junction with no significant effect of the high packet error rate. Fig. 4c, 4d, 4e, and 4f show that the travel rate decreases significantly when there are more CAVs present on road and the communication is reliable. The improvement in traffic efficiency is, however, not significant in the presence of imperfect communication compared to the baseline scenario i.e. only HDVs. The results indicate that a high penetration rate of CAVs corresponds to higher improvement in traffic efficiency, as expected. These results also show that the travel rate at 70% CAV penetration with packet error rate of 0.7 is actually a little bit worse than as compared to performance at 40% without packet drops.

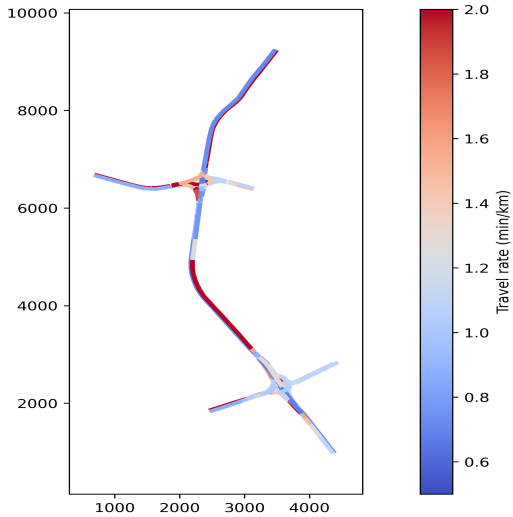

Fig. 2: Travel rate at 0% penetration rate

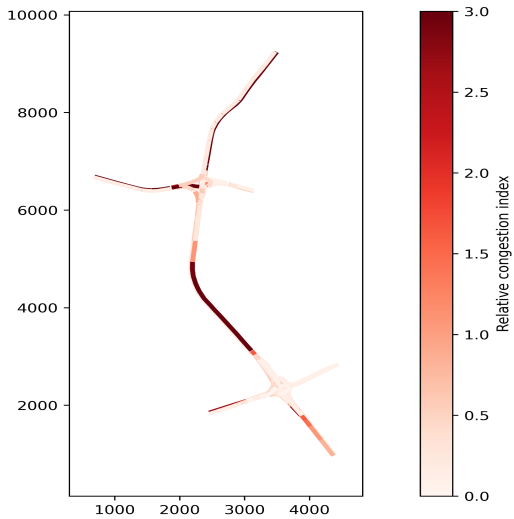

Fig. 3: Relative congestion index at 0% penetration rate

Fig. 5 shows the simulation results of relative congestion index at different penetration rates (20%, 40%, 70%) with and without packet drops. From Fig. 5a and 5b, it is observed that at low penetration rate i.e. 20%, CAVs increase the road traffic congestion while no significant effect of the high packet error rate is observed. Fig. 5c and 5e show that the traffic congestion reduces significantly when there are more CAVs present on road i.e. at high penetration rates, however, the traffic flow conditions worsen significantly in the presence of imperfect communication as shown in Fig. 5d and 5f.

The results of travel rate and relative congestion index at CAVs different penetration rates with and without packet drops are summarised in table V. This table shows that CAVs can reduce the travel rate and traffic congestion significantly at high penetration rates, however, a diminishing effect is observed in imperfect communication.

TABLE V: Summary of traffic efficiency results

| MPR (%) | PER | Travel rate (min/km) | RCI |
| --- | --- | --- | --- |
| 0 | NA | 1.5208 | 0.9712 |
| 20 | 0 | 1.5416 | 0.9825 |
| 20 | 0.7 | 1.5450 | 0.9891 |
| 40 | 0 | 1.3998 | 0.7977 |
| 40 | 0.7 | 1.4643 | 0.8925 |
| 70 | 0 | 1.3597 | 0.7330 |
| 70 | 0.7 | 1.5034 | 0.9435 |

## V. DATA ANALYSIS AND CHALLENGES

As detailed in Section IV, a total of seven experiments were run in parallel, producing over 200GB of data in the OMNeT++ vector format (*.vec). Although SUMO can be configured to provide useful output for each edge of the simulated road network, it was not found possible to override the path of this output while running multiple simulations in parallel. Therefore, the data needed to be emitted in a different format, and reconstructed afterwards [21]. Initially, the scavetool program supplied with OMNeT++ was used to export the data to CSV format [14]. This proved to be untenable, due to both the size of the resulting files and the time taken to perform the conversion. To work around this, we used Python scripts created by the author in [21] to convert a vector file sequentially into a SQLite database and then reconstruct the edge-based vehicle data that SUMO is capable of analysing using available visualization tools.

Performing road traffic simulations in realistic scenarios is found to be extremely resource extensive. Simulation for the congested traffic hours at 70% penetration rate took over 72 hours of real time to complete. Additionally, an enormous volume of data i.e. over 39.5 million rows of data is produced by simulations of such large-scale networks, hence, the data analysis was a daunting task. Finally, due to the single-threaded nature of both the SUMO and OMNeT++, multiple simulations needed to be run in parallel [21]. To support parallel simulation execution, simulations were performed at the personal desktop computer and high performance computing cluster.

## VI. CONCLUSIONS AND FUTURE SCOPE

This paper investigated the impact of CAVs on mixed traffic efficiency in congested traffic scenarios considering

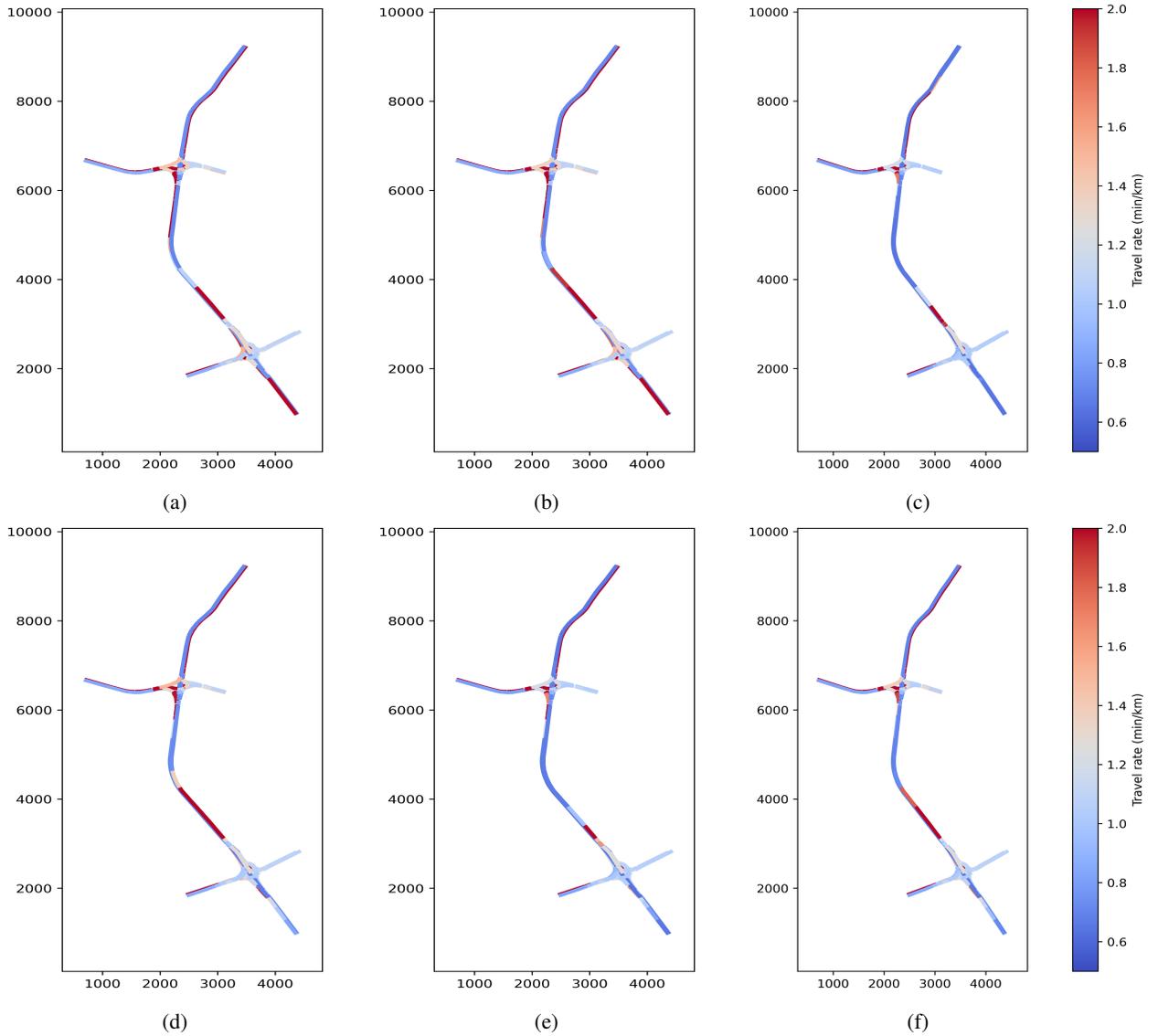

Fig. 4: Travel rate (min/km) at different penetration rates, with and without packet drops, (a) MPR 20% and PER 0%, (b) MPR 20% and PER 70%, (c) MPR 40% and PER 0%, (d) MPR 40% and PER 70%, (e) MPR 70% and PER 0%, (f) MPR 70% and PER 70%

imperfect communication environments and large-scale road network. Simulation results showed that high penetration rates of CAVs provide significant improvement in traffic performance. The travel rate and relative congestion index measured at different penetration rates and packet error rates indicate that CAVs can improve traffic efficiency by employing the CACC technology in their car-following control. However, in imperfect communication, traffic congestion increase drastically at high penetration rates (0.67% increase at 20% MPR, 28.71% increase at 40% MPR, 11.88% increase at 70% MPR) resulting in reduced traffic efficiency. In summary, CAVs with reliable communication provide improvement in traffic efficiency as the penetration rate increases, however, this improvement is not very significant in the presence of packet losses.

To the best of our knowledge, this paper is the first study that investigated the impact of CAVs on traffic performance in imperfect communication for a large-scale road network. There are a few interesting points which would be considered as the future scope of this paper. While this paper employs a CACC controller which was designed based on information obtained from the preceding vehicle only, an interesting future work direction would be to investigate traffic performance by employing a CACC controller in which an ego vehicle may obtain information from multiple leading vehicles [20]. Furthermore, in this paper, when a CAV cannot establish a communication link with its leading vehicle, its car-following mode fall-backs to ACC, resulting in reduced traffic efficiency, however, future work may consider designing a robust control technique which will

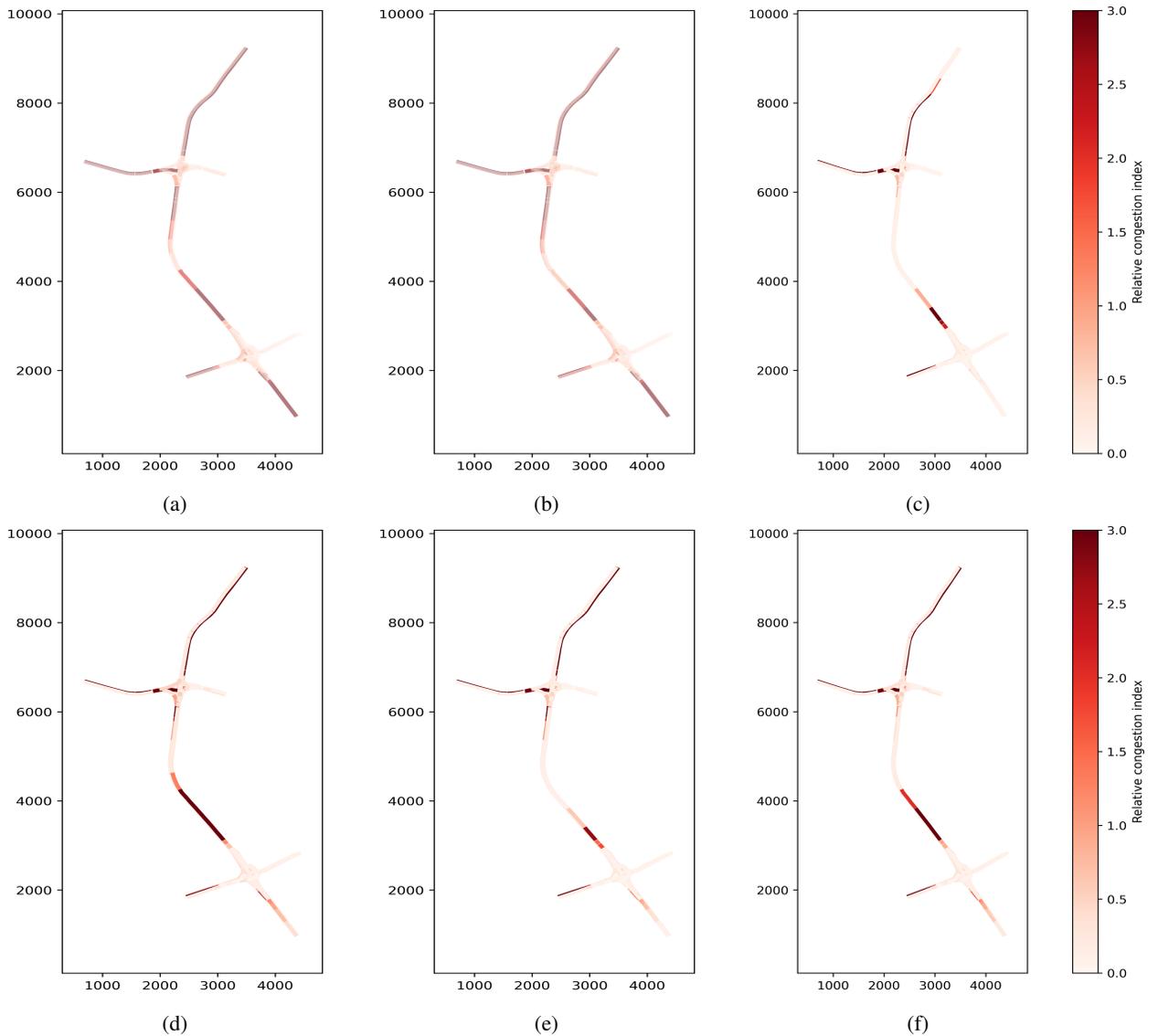

Fig. 5: Relative congestion index at different penetration rates, with and without packet drops, (a) MPR 20% and PER 0%, (b) MPR 20% and PER 70%, (c) MPR 40% and PER 0%, (d) MPR 40% and PER 70%, (e) MPR 70% and PER 0%, (f) MPR 70% and PER 70%

decrease the detrimental effects of imperfect communication on CAV performance [16].

## ACKNOWLEDGMENTS

This research was supported by a Trinity College Provost PhD award, generously funded through alumni donations and Trinity College Dublin's Commercial Revenue Unit, and a research grant from Science Foundation Ireland (SFI) under Grant Number 16/SP/3804 (Enable).

The authors would also like to thank Dr. Maxime Gueriau for his generous help with setting up the experiments for this work.